\documentstyle[prl,aps,epsbox]{revtex}
\makeatletter
\begin{document}
\twocolumn
\newcommand{\bdx}{\mbox{\boldmath $x$}}
\newcommand{\cL}{{\cal{L}}}
\newcommand{\cH}{{\cal{H}}}
\newcommand{\cV}{{\cal{V}}}
\newcommand{\tilphi}{\tilde{\phi}}
\newcommand{\tilvarphi}{\tilde{\varphi}}
\newcommand{\tilpi}{\tilde{\pi}}
\newcommand{\phizero}{\phi_0}
\newcommand{\varphizero}{\varphi_0}
\newcommand{\an}{a_n}
\newcommand{\andg}{a_n^\dagger}
\newcommand{\pn}{p_n^+}
\newcommand{\xp}{x^+}
\newcommand{\xm}{x^-}
\newcommand{\intL}{\int_{-L}^L}
\newcommand{\mue}{\mu_{\rm e}}
\newcommand{\sn}{\mbox{\rm sn}}
\newcommand{\cn}{\mbox{\rm cn}}
\newcommand{\dn}{\mbox{\rm dn}}
\newcommand{\gd}{\mbox{\rm gd}}
\title{Manifestation of a nontrivial vacuum 
in discrete light cone quantization}
\author{Takanori Sugihara and Masa-aki Taniguchi}
\address{Department of Physics, 
Nagoya University, Chikusa, Nagoya 464-8602, Japan}
\maketitle

\begin{abstract}
We study a (1+1)-dimensional $\lambda\phi^4$ model 
with a light-cone zero mode and constant external source 
to describe spontaneous symmetry breaking. 
In the broken phase, we find degenerate vacua 
and discuss their stability based on effective-potential analysis. 
The vacuum triviality is spurious in the broken phase 
because these states have lower energy than Fock vacuum. 
Our results are based on the variational principle. 
\end{abstract}

\pacs{PACS numbers: 11.10.Gh, 11.10.Kk, 11.30.Qc}

Light-cone quantization \cite{dirac} has been studied 
to clarify nonperturbative aspects of field theories \cite{lf} 
and used to provide nonperturbative formulation of 
M-theory \cite{m}. 
This framework simplifies dynamics of quantum field theories 
since it prohibits vacuum diagrams kinematically \cite{lks}. 
It has also a possibility of calculating wavefunctions 
of physical states in a nonperturbative manner. 
Because of these preferable properties, 
we hope for this framework playing 
a complementary role to lattice theories.

DLCQ (discrete light cone quantization) 
is the most well-defined treatment of light-cone quantization, 
which enables clear separation of zero mode \cite{my,casher,pb}. 
It has been believed that the true vacuum is trivial and 
only the zero mode is responsible for 
spontaneous symmetry breaking (SSB) 
in scalar field theories \cite{my,kty}. 
There are some studies that rely on a combination 
of constrained zero modes and trivial vacuum \cite{zero}. 
The vacuum triviality, however, results from an assumption 
that normal-ordered Hamiltonians are positive semidefinite. 
In this letter, we show that this assumption is not always true. 
We include a zero mode and external source 
in analytic variational calculations to show the existence 
of nontrivial vacuum with lower energy than trivial Fock vacuum. 
Our results are based on previous works on 
zero-mode singularity \cite{kty} and 
quantum solitons \cite{sugi,rt} in DLCQ. 
We examine a possibility that zero modes reproduce SSB, 
which has not been considered in Ref. \cite{sugi,rt}. 

In order to define an effective potential, we consider 
the generating functional of Green's functions, $W[J]=-i\ln Z[J]$. 
We can express it using a Hamiltonian $H[J]$ 
when the external source is time-independent $J(x)=J(\bdx)$ 
\cite{sugi,coleman}, 
\[
        H[J]|0_J\rangle=w[J]|0_J\rangle, \quad
        H[J]=H-\int_{-L}^L d\bdx J(\bdx)\phi(\bdx), 
\]
where $w[J]=-W[J]/T$ is energy of the ground state $|0_J\rangle$. 
$T$ is the time difference between the initial and final states. 
If we are just interested in the ground state (vacuum), 
we can consider an effective potential 
to reduce the problem into simpler one. 
It is defined as a Legendre transform 
$V(\varphizero)\equiv w(J)+J\varphizero
=\langle0_J|H|0_J\rangle/2L$ 
with a constant source $J(x)=J$ and vacuum-energy density $w(J)=w[J]/2L$. 
A vacuum expectation value (VEV) of a zero mode is given by 
$\varphizero=-dw(J)/dJ=\langle 0_J|\phizero|0_J \rangle$. 
We can calculate the effective potential $V(\varphizero)$ 
if energy and wavefunction of the ground state are known. 
Our purpose is to identify the true vacuum 
in the (1+1)-dimensional $\lambda\phi^4$ model 
with a double-well classical potential. 
We perform variational calculations for 
the following normal-ordered Hamiltonian. 
\begin{equation}
    H(J)=\intL dx^-
    \left(
    -\frac{\mu^2}{2}:\phi^2:+\frac{\lambda}{4!}:\phi^4:
    -J:\phi:
    \right). 
    \label{hamiltonian}
\end{equation}
Hereafter, we will designate the space coordinate $\xm$ by $x$. 
The field operator is decomposed into 
zero and nonzero modes $\phi(x)=\phizero+\tilphi(x)$, 
where $(O)_0 \equiv \intL dx O(x)/2L$. 
In DLCQ with periodic boundary conditions, 
the zero mode $\phizero$ is constrained \cite{my}, 
\begin{equation}
    \Phi[\tilphi]=
    -\mu^2:\phizero: + \frac{\lambda}{6}:(\phi^3)_0:-J=0. 
    \label{zmcq}
\end{equation}
Operator ordering has been chosen so that it satisfies 
$dw(J)/dJ=-\varphizero$. 
It is the consistency condition to be satisfied 
between the Hamiltonian (\ref{hamiltonian}) and 
zero-mode constraint (\ref{zmcq}). 
We will discuss it later in Eq. (\ref{st}). 
The zero mode is an operator functional of the nonzero mode 
\begin{equation}
    \tilphi(x)=
    \sum_{n=1}^\infty\frac{1}{\sqrt{4\pi n}}
        \left(
        \an e^{-i\pn x}+\andg e^{i\pn x}
        \right), 
\end{equation}
where $[a_m,a_n^\dagger]=\delta_{m,n}$ and $a_m|0_{\rm F}\rangle=0$. 
We truncate the system by introducing one-coherent state 
$|\tilvarphi\rangle=e^{\sum_{n=1}^\infty
(\tilvarphi_n \andg -\frac{1}{2}\tilvarphi_n^*\tilvarphi_n)}
|0_{\rm F}\rangle$. 
It provides useful formulas 
$a_n|\tilvarphi\rangle=\tilvarphi_n|\tilvarphi\rangle$ 
and 
$\langle\tilvarphi|:O[\tilphi]:|\tilvarphi\rangle=O[\tilvarphi]$, 
where 
$\varphi(x)\equiv\langle\tilvarphi|\phi(x)|\tilvarphi\rangle
=\varphizero+\tilvarphi(x)$. 
Our variational parameter is $\tilvarphi(x)$. 

\begin{figure}[h]
  \begin{center}
    \epsfile{file=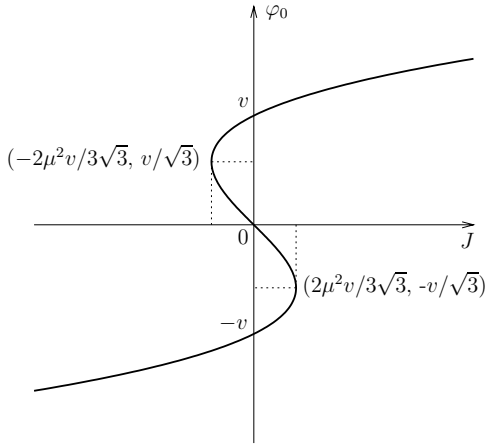,scale=0.7}
  \end{center}
  \caption{Functional relationship (\ref{zmc}) 
between $\varphizero$ and $J$ is shown 
when Fock space is truncated 
with the one-coherent approximation. 
The region $|\varphizero|<v/\sqrt{3}$ is unstable 
(see discussions given below Eq. (\ref{d2})). 
When $|\varphizero|\ge v$, 
zero mode $\varphizero$ increases monotonously, $d\varphizero/dJ>0$. 
    }
  \label{fig1}
\end{figure}

We take the continuum limit with $P^+=\pi M/L$ fixed, 
where $M$ is called the harmonic resolution. 
We include this constraint on $P^+$ in Hamiltonian 
using the Lagrange's undetermined multiplier. 
\begin{eqnarray}
    &&E\equiv
    \langle \tilvarphi|H_\beta (J)|\tilvarphi \rangle
    \label{energy}
    \\
    =&&\intL dx
    \left[
      \beta\left( \left(\frac{d\tilvarphi}{dx}\right)^2
      -\frac{\pi M}{2L^2}\right)
      -\frac{\mu^2}{2}\varphi^2
      +\frac{\lambda}{4!}\varphi^4
      -J\varphi 
      \right],
    \nonumber
\end{eqnarray}
where $w(J)=E/2L$. 
Equation (\ref{zmcq}) and the stationary condition 
for (\ref{energy}) give the following 
coupled equations for the zero and nonzero modes. 
\begin{eqnarray}
    &
    \displaystyle
    -\mu^2\varphizero+\frac{\lambda}{6}\varphizero^3=J, 
    &
    \label{zmc} 
    \\
    &
    \displaystyle
    -2\beta\frac{d^2\tilvarphi}{dx^2}
    +\mue^2\tilvarphi
    +\frac{\lambda}{2}\varphizero\tilvarphi^2
    +\frac{\lambda}{6}\tilvarphi^3=0, 
    &
    \label{eom}
\end{eqnarray}
where $\mue^2\equiv -\mu^2 + \lambda\varphizero^2/2$. 
We have the following zero-mode solutions 
to (\ref{zmc}) when $J=0$ (see Fig. \ref{fig1}). 
\begin{equation}
    \varphizero=0, \pm v, \quad 
    v\equiv \sqrt{\frac{6\mu^2}{\lambda}}. 
\end{equation}
The system defined in a finite box can describe SSB 
if the vanishing $J$ limit is taken after all calculations. 
We should choose $\varphizero=\pm v$ 
but not $\varphizero=0$ as physical solutions to (\ref{zmc})
when $J=0$, 
because the solution $\varphizero=0$ is not connected 
continuously to expectation values for large $J\ne 0$. 
The region $|\varphizero|< v/\sqrt{3}$ is unphysical. 

If one attempts to see vacuum physics 
without introducing an external source, 
wrong solutions may be obtained. 
Symmetric phase would be safe without an external source, 
but broken phase is not. 
We will explain how the true vacuum solution 
appears by paying attention to the effects of 
the zero mode and external source on the ground state. 
We will also discuss vacuum stability in Eq. (\ref{d2}) 
based on the effective potential. 

Solutions to Eq. (\ref{eom}) 
must satisfy the following two conditions simultaneously: 
\begin{itemize}
\item[(i)] The solution $\tilvarphi$ to (\ref{eom}) 
and the $n$-th derivative 
$\tilvarphi^{(n)}=d^n\tilvarphi/dx^n$ 
must be periodic at the boundaries $x=\pm L$: 
$\tilvarphi^{(n)}(-L)=\tilvarphi^{(n)}(L), \quad
n=0,1,2,...$ 
\item[(ii)] The solution $\tilvarphi$ to (\ref{eom}) 
must be nonzero mode: 
$(\tilvarphi(x))_0=\intL dx \tilvarphi(x)/2L=0$. 
\end{itemize}

For convenience, let us regard $x$ as time and
consider the following dummy Hamiltonian $\cH$ 
that reproduces the equation of motion (\ref{eom}). 
\begin{equation}
    \cH=\frac{1}{2}(\partial\tilde{\varphi})^2+\cV(\tilvarphi), 
\end{equation}
where $\cV$ is a dummy potential 
\begin{equation}
    \cV(\tilvarphi)=
    -\frac{1}{2\beta}
    \left(
      \frac{\mue^2}{2}\tilvarphi^2
      +\frac{\lambda}{6}\varphizero\tilvarphi^3
      +\frac{\lambda}{4!}\tilvarphi^4
      -\frac{\mu^2}{2}\varphizero^2
      +\frac{\lambda}{4!}\varphizero^4
    \right). 
    \label{dumpot}
\end{equation}

\begin{figure}[h]
  \begin{center}
    \epsfile{file=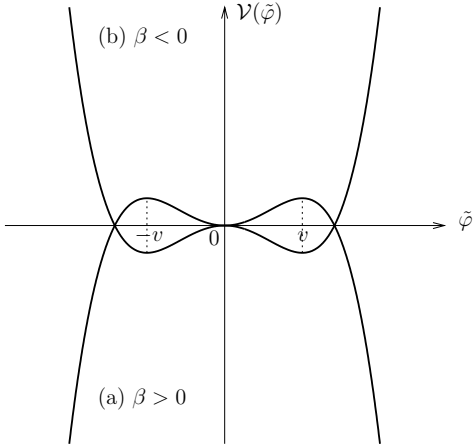,scale=0.7}
  \end{center}
  \caption{The dummy potential ${\cal V}$ is drawn as a function 
of $\tilvarphi$ when $\varphizero=0$: (a) positive $\beta$ and 
(b) negative $\beta$. 
    }
  \label{fig2}
\end{figure}

We first solve (\ref{eom}) especially when $J=0$ and $\varphizero=0$. 
As mentioned before, this gives physically unacceptable solutions. 
However, these solutions are technically helpful for the purpose of 
calculating energy and wavefunction of the true degenerate vacua 
with nonzero VEVs as shown later. 
When $J=0$ and $\varphizero=0$, Eq. (\ref{eom}) reduces to 
\begin{equation}
        \frac{d^2\tilvarphi}{dx^2}=
        \frac{1}{2\beta}
        \left(
        -\mu^2\tilvarphi
        +\frac{\lambda}{6}\tilvarphi^3
        \right), 
        \label{eom2}
\end{equation}
and the dummy potential $\cV$ is 
\begin{equation}
    \cV(\tilvarphi)=
    -\frac{1}{2\beta}
    \left(
      -\frac{\mu^2}{2}\tilvarphi^2+\frac{\lambda}{4!}\tilvarphi^4
    \right). 
    \label{dumpot1}
\end{equation}
When the parameter $\beta$ is positive, the dummy potential 
$\cV(\tilvarphi)$ is bounded from above (see Fig. \ref{fig2}(a)). 
From the condition (i), 
motion of a particle must be periodic. 
Namely, the particle must reside between the two maximums 
of the dummy potential. 
If the condition (i) is satisfied, 
the condition (ii) is also satisfied 
since the particle oscillates around the origin 
in the symmetric dummy potential. 
The solution to (\ref{eom2}) is 
\begin{equation}
    \tilvarphi_{\rm sn}(x)
    =\left[\frac{12\mu^2k^2}{\lambda(k^2+1)}\right]^{1/2}
    \sn(a_{\rm sn}x,k), 
    \label{sol1}
\end{equation}
where $\sn$ is a Jacobian elliptic function and 
$0\le k\le 1$ \cite{rt}. 
The values of the parameters $a_{\rm sn}$ and $k$ are determined 
so that the solution (\ref{sol1}) satisfies 
both the conditions (i) and (ii); 
$k=0$ gives $\tilvarphi(x)=0$ and $M=0$, which correspond to 
trivial Fock vacuum with $\varphizero=0$ and $P^-=0$. 
$k=1$ is not acceptable since it gives 
$\tilvarphi\sim\tanh(ax)$, 
which is an odd function and 
cannot satisfy the periodicity condition (i). 
When $0<k<1$, we have $a_{\rm sn}=2NK(k)/L$ 
($N$ is a natural number and $K(k)$ is the complete 
elliptic integral of the first kind) 
from the periodicity condition (i) and 
it satisfies the condition (ii).   
In this case, the solution (\ref{sol1}) is acceptable 
and gives the following energy and harmonic resolution. 
\begin{eqnarray}
    &
    \displaystyle
    \frac{E_{\rm sn}}{2L}=-\frac{6\mu^4 k^2}{\lambda(k^2+1)^2}
        +\frac{576N^2\mu^6 k^4 I_{\rm sn}^2(k)}
        {\pi\lambda^2(k^2+1)^3 M}, 
    &
    \label{e1}
    \\
    &
    \displaystyle
    M
    =\frac{96N\mu^2 k^2 I_{\rm sn}(k)K(k)}{\pi\lambda(k^2+1)}, 
    &
    \\
    &
    \displaystyle
    I_{\rm sn}(k)\equiv \int_0^1 df \sqrt{(1-f^2)(1-k^2f^2)}. 
    &
\end{eqnarray}
In the limit $k\to 1$ with $N=1$, the energy (\ref{e1}) takes 
the minimum value, and the harmonic resolution goes to infinity 
since $K(k)$ diverges at $k=1$ giving the continuum limit $M\to\infty$. 
This is the solution given in Ref. \cite{rt}. 
When the parameter $\beta$ is negative, 
the condition (i) is automatically satisfied 
since the dummy potential ${\cal V}(\tilvarphi)$ is bounded from below 
and a particle oscillates necessarily with a fixed period 
(see Fig. \ref{fig2}(b)). 
Equation (\ref{eom2}) has two types of solutions in this case. 
We can also express them using Jacobian elliptic functions 
$\cn(x,k)$ and $\dn(x,k)$. 
When a particle oscillates around the origin, 
a solution is 
\begin{equation}
    \tilvarphi_{\rm cn}(x)
    =\left[\frac{12\mu^2k^2}{\lambda(2k^2-1)}\right]^{1/2}
    \cn(a_{\rm cn}x,k). 
    \label{sol2}
\end{equation}
The solution (\ref{sol2}) is acceptable as an exited state 
since it satisfies the conditions (i) and (ii) simultaneously 
with higher energy than (\ref{sol1}). 
When a particle oscillates around one of the minimums 
of the dummy potential, solutions are 
\begin{equation}
    \displaystyle
    \tilvarphi_{\rm dn}(x)
    =\pm\left[\frac{12\mu^2}{\lambda(2-k^2)}\right]^{1/2}
    \dn(a_{\rm dn}x,k). 
    \label{sol3}
\end{equation}
The solutions (\ref{sol3}) are not acceptable 
since they cannot satisfy the condition (ii). 

When $J=0$ and $\varphizero=0$, 
the candidate solution for the ground state 
is (\ref{sol1}) with $a_{\rm sn}=2NK(k)/L$ and $k\to 1$ 
derived for positive $\beta$ 
since it gives the lowest energy. 
However, we discard the solution (\ref{sol1}) 
since the state formed by it cannot be 
connected to solutions for non-zero external source $J\ne 0$. 
In addition, mass squared goes to negative infinity 
$P^\mu P_\mu\to-\infty$ in the continuum limit $M\to\infty$ 
since the first term of the energy (\ref{e1}) is nonzero and 
negative. 
This is the reason why all past calculations 
have not been successful 
in calculating mass squared stably 
in the broken phase \cite{sugi}. 

When the limit $J\to 0$ is taken starting from 
a sufficiently large $J$, 
an effective potential chooses one of $\varphizero=\pm v$ 
depending on the sign of $J$. 
Nonzero values of $J$ resolve the degeneracy of the two vacua, 
which restores in the limit $J\to 0$. 
In this case, 
the conditions (i) and (ii) select $\dn$-type oscillation 
around the minimum $\tilvarphi=0$ of the dummy potential. 
The situation is completely different 
from the $\varphizero=0$ case. 
We obtain the following dummy potential by substituting 
$\varphizero=v$ into (\ref{dumpot}) (it is enough to consider 
one of the two degenerate vacua $\varphizero=\pm v$). 
\begin{equation}
    \cV(\tilvarphi)=
    -\frac{1}{2\beta}
    \left(
      \mu^2\tilvarphi^2
      +\frac{\lambda}{6}v\tilvarphi^3
      +\frac{\lambda}{4!}\tilvarphi^4
      -\frac{3\mu^4}{2\lambda}
    \right). 
    \label{dumpot2}
\end{equation}
This is the case when the nonzero mode is shifted with 
$\tilvarphi\to\tilvarphi+v$ in (\ref{dumpot1}). 
Therefore, we obtain four possible oscillations 
by shifting the solutions 
(\ref{sol1}), (\ref{sol2}), and (\ref{sol3}) 
with $\tilvarphi\to\tilvarphi-v$. 
The first is $\sn$-type oscillation around $\tilvarphi=-v$. 
The second is $\cn$-type oscillation around $\tilvarphi=-v$. 
The third is $\dn$-type oscillation around $\tilvarphi=-2v$. 
However, all of them are unacceptable 
since they cannot satisfy the condition (ii). 
The following $\dn$-type solution is 
a physically acceptable oscillation: 
\begin{equation}
    \tilvarphi(x)
    =\left[\frac{12\mu^2}{\lambda(2-k^2)}\right]^{1/2}\dn(ax,k)-v, 
    \label{sol}
\end{equation}
which oscillates around the origin $\tilvarphi=0$ and can satisfy 
the conditions (i) and (ii) simultaneously. 
Its energy and total momentum are 
\begin{eqnarray}
    \frac{E}{2L}&=&
     \frac{6\mu^4(k^2-1)}{\lambda(2-k^2)^2}
    -\frac{144\mu^6 I_{\rm dn}^2(k)}{\pi\lambda^2(2-k^2)^3 M}, 
    \label{v}
    \\
    P^+&=&\frac{\pi M}{L}
    =\frac{24\mu^2 I_{\rm dn}(k)}{\lambda(2-k^2)}a_, 
    \label{p}
    \\
    I_{\rm dn}(k)&\equiv& \int_{{\rm dn}(aL,k)}^1 
    df \sqrt{(1-f^2)(f^2-1+k^2)}, 
    \nonumber
\end{eqnarray}
where $0\le k\le 1$; 
$k=0$ is not acceptable since $\tilvarphi(x)=0$ gives 
$\varphizero=\langle 0_{\rm F}|:\phizero:|0_{\rm F}\rangle=0$ 
that contradicts $\varphizero=v$. 
When $0<k<1$, there exists no value of the parameter 
$a=NK(k)/L$ that can satisfy the condition (ii). 
When $k=1$, we have 
\begin{equation}
    \tilvarphi(x)
    =v
    \left[
      \frac{\sqrt{2}}{\cosh(ax)}-1
    \right]. 
    \label{sol_k=1}
\end{equation}
The condition (ii) requires the following relation to hold. 
\begin{equation}
    \sqrt{2}\gd(aL)-aL=0, 
\end{equation}
where $\gd$ is the Gudermann function. 
This has a solution $aL\sim 1.72$. 
In the continuum limit $L\to\infty$, we have $a\to 0$ 
which gives zero total momentum $P^+=0$. 
Since the energy (\ref{v}) is negative and 
lower than Fock vacuum in the continuum limit, 
we identify it as one of the true vacua. 

In order to check the consistency of the operator ordering 
between Hamiltonian (\ref{hamiltonian}) and 
zero-mode constraint (\ref{zmcq}), 
we examine the first derivative of $w(J)$ 
with respect to $J$ using (\ref{zmc}) and (\ref{eom}). 
\begin{eqnarray}
    &&\frac{dw}{dJ}+\varphizero
    \nonumber
    \\
    &=&\intL \frac{dx}{2L}
    \left[
      \frac{d\tilvarphi}{dJ}
      \left(
        \mue^2\tilvarphi
        +\frac{\lambda}{2}\varphizero\tilvarphi^2
        +\frac{\lambda}{6}\tilvarphi^3
      \right)
      +\frac{d\varphizero}{dJ}\Phi[\tilvarphi]
    \right]
    \nonumber
    \\
    &=&
    -\beta\frac{d}{dJ}\left(\frac{P^+}{2L}\right). 
    \label{st}
\end{eqnarray}
Since the parameters $P^+$ and $L$ are given by hand 
independent of $J$, we obtain the desired consistency condition 
$dw/dJ=-\varphizero$. 
We can use this relation to discuss vacuum stability also. 
We have the following relation 
from the definition of the effective potential 
$V(\varphi_0)=w(J)+J\varphi_0$. 
\begin{equation}
  \frac{d^2V}{d\varphi_0^2}=\frac{dJ}{d\varphi_0}. 
  \label{d2}
\end{equation}
The state given by (\ref{sol_k=1}) is stable and hence 
can be regarded as one of the true vacua since (\ref{d2}) 
is positive when $\varphi_0 = v$. 
On the other hand, the state given by (\ref{sol1}) is unstable 
since (\ref{d2}) is negative when $\varphi_0=0$. 
In $-v/\sqrt{3}<\varphizero<v/\sqrt{3}$, 
energy decreases as $J$ increases. 
If $\phizero$ or $J$ is not introduced, 
one cannot observe this instability of the state (\ref{sol1}) 
since $V(\varphizero)$ and hence (\ref{d2}) are not available. 

We conclude that there exist nontrivial degenerate vacua 
other than Fock vacuum 
in the (1+1)-dimensional $\lambda\phi^4$ model with 
a double-well classical potential. 
We have shown stability of the obtained vacua 
based on the effective-potential analysis.  
The essential point of our analysis is introduction of 
a zero mode and external source. 

In general, there are singlet and non-singlet sectors 
of $Z_2$ symmetry. 
One-coherent state $|\tilvarphi\rangle$ is a mixed state of 
both the sectors. 
We have shown that there exist non-singlet vacua 
with lower energy than Fock vacuum 
when the classical potential has a double-well shape. 
The mixing of the singlet and non-singlet sectors is 
a consequence of introduction of 
an explicitly symmetry-breaking interaction $J\phi_0$ 
in the Hamiltonian (\ref{hamiltonian}). 

The issues of critical exponents remain still open 
until quantitatively reliable calculations are done. 
We should perform variational calculations without 
assuming vacuum triviality also in the case 
when a classical potential is convex. 
However, we will need to interpret the difference among 
normal (ours) and other operator orderings (such as Weyl ordering) 
concerning to the origin of VEVs of zero modes. 
When the Hamiltonian and zero-mode constraint are normal-ordered, 
trivial Fock vacuum just gives 
$\langle 0_{\rm F}|:H:|0_{\rm F}\rangle=0$ and 
$\langle 0_{\rm F}|:\phizero:|0_{\rm F}\rangle =0$ 
(i.e. there is no SSB if vacuum triviality is assumed 
for normal-ordered operators). 

Finally, we point out the importance of small 
momentum components near the zero mode. 
The reason why the true-vacuum solution (\ref{sol_k=1}) 
gives $P^+=0$ in the continuum limit $L\to \infty$ 
is that its slowly-changing configuration 
is mainly composed of small momenta. 
This is an extended description of the accumulating point 
discussed before in Ref. \cite{ny}. 
On the other hand, the solution (\ref{sol1}) with 
$a_{\rm sn}=2NK(k)/L$ and $k\to 1$ needs 
large momentum components to describe its singular behavior 
at $x=0$ and $\pm L$, 
which gives infinite harmonic resolution $M\to\infty$. 

We would like to express sincere thanks to K. Yamawaki 
for valuable discussions. 
T.S. would like to thank K. Harada and K. Yazaki 
for helpful comments. 
T.S. was supported by the Grant-in-Aid for Scientific 
Research Fellowship, No. 11000480.


\vspace{-0.5cm}

\begin{thebibliography}{30}
%
\bibitem{dirac} P.A.M. Dirac, 
Rev. Mod. Phys. {\bf 21}, 392 (1949). 
\bibitem{lf} K.G. Wilson, T.S. Walhout, A. Harindranath, 
W. M. Zhang, R.J. Perry, and S.D. Glazek, 
Phys. Rev. D {\bf 49}, 6720 (1994); 
K. Yamawaki, 
``{\it New Nonperturbative Methods and Quantization 
on the Light Cone}'', 
Les Houches, France, Feb. 24-Mar. 7, 1997; 
S.J. Brodsky, H.C. Pauli, and S.S. Pinsky, 
Phys. Rept. 301, 299-486 (1998). 
\bibitem{m} T. Banks, W. Fischler, S. H. Shenker, and L. Susskind, 
Phys. Rev. D {\bf 55}, 5112 (1997);
L. Susskind, hep-th/9704080. 
\bibitem{lks} H. Leutwyler, J.R. Klauder, and L. Streit, 
Nuovo Cim. {\bf 66A}, 536 (1970). 
\bibitem{my} T. Maskawa and K. Yamawaki, 
Prog. Theor. Phys. {\bf 56}, 270 (1976). 
\bibitem{casher} A. Casher, 
Phys. Rev. D {\bf 14}, 452 (1976). 
\bibitem{pb} H. C. Pauli and S. J. Brodsky, 
Phys. Rev. D {\bf 32}, 1993 (1985); {\it ibid} 2001 (1985); 
T. Eller, H.C. Pauli, and S.J. Brodsky,
Phys. Rev. D {\bf 35}, 1493 (1987); 
T. Eller and H.C. Pauli, 
Z. Phys. C {\bf 42}, 59 (1989)
\bibitem{kty} Y. Kim, S. Tsujimaru, and K. Yamawaki, 
Phys. Rev. Lett. {\bf 74}, 4771 (1995);
S. Tsujimaru and K. Yamawaki, 
Phys. Rev. D {\bf 57}, 4942 (1998). 
\bibitem{zero} 
T. Heinzl, C. Stern, E. Werner and B. Zellermann,
Z. Phys. C {\bf 72}, 353 (1996); 
D.G. Robertson, 
Phys. Rev. D {\bf 47}, 2549 (1993); 
C.M. Bender, S.S. Pinsky, and B. van de Sande, 
Phys. Rev. D {\bf 48}, 816 (1993); 
S.S. Pinsky and B. van de Sande, 
Phys. Rev. D {\bf 49}, 2001 (1994); 
S.S. Pinsky, B. van de Sande, and J.R. Hiller, 
Phys. Rev. D {\bf 51}, 726 (1995). 
\bibitem{sugi} T. Sugihara, 
Phys. Rev. D {\bf 57}, 7373 (1998).
\bibitem{rt} J. S. Rozowsky and C. B. Thorn, 
Phys. Rev. Lett. {\bf 85}, 1614 (2000). 
\bibitem{coleman} 
S. Coleman, {\it Aspects of Symmetry} 
(Cambridge University Press, Cambridge, England, 1985). 
\bibitem{ny} N. Nakanishi and K. Yamawaki, 
Nucl. Phys. B {\bf 122}, 15 (1977). 
\end{thebibliography}
\end{document}